\documentclass[prb, twocolumn,superscriptaddress,preprintnumbers,a4paper,amsmath,amssymb,showpacs,floatfix]{revtex4}

\usepackage{graphicx}
\usepackage{color}\color[rgb]{0.000,0.000,0.000} 
\usepackage{amssymb} 
\usepackage{amsmath} 

\begin{document}

\newcommand{\co}{CoV$_2$O$_6$}
\newcommand{\mn}{MnV$_2$O$_6$}
\newcommand{\nb}{CoNb$_2$O$_6$}
\newcommand{\ca}{Ca$_3$Co$_2$O$_6$}

\newcommand{\te}{$_{t}$}
\newcommand{\ot}{$_{o}$}
\newcommand{\Tc}{T$_{C}$}
\newcommand{\Ts}{T$_{s}$}
\newcommand{\Tn}{$T_\mathrm{N}$}
\newcommand{\MuB}{$\mu_\mathrm{B}$}

\title{Metamagnetism and soliton excitations in the modulated ferromagnetic Ising chain \co}
\author{Simon A. J. Kimber}
\email[Email of corresponding author:]{kimber@esrf.fr}
\affiliation{European Synchrotron Radiation Facility (ESRF), 6 rue Jules Horowitz, BP 220, 38043  Grenoble Cedex 9, France}
\affiliation{Helmholtz-Zentrum Berlin f\"ur Materialien und Energie (HZB),  Hahn-Meitner Platz 1, 14109, Berlin, Germany}
\affiliation{School of Chemistry, Joseph Black Building, King's Buildings, West Mains Road, Edinburgh, EH9 3JJ}


\author{Hannu Mutka}

\affiliation{Institut Max von Laue-Paul Langevin, 6 rue Jules Horowitz, BP 156, F-38042, Grenoble Cedex 9, France}

\author{Tapan Chatterji}

\affiliation{Institut Max von Laue-Paul Langevin, 6 rue Jules Horowitz, BP 156, F-38042, Grenoble Cedex 9, France}

\author{Tommy Hofmann}
\affiliation{Helmholtz-Zentrum Berlin f\"ur Materialien und Energie (HZB),  Hahn-Meitner Platz 1, 14109, Berlin, Germany}

\author{Paul. F. Henry}
\affiliation{Helmholtz-Zentrum Berlin f\"ur Materialien und Energie (HZB),  Hahn-Meitner Platz 1, 14109, Berlin, Germany}
\affiliation{European Spallation Source ESS AB, Box 176, 22100, Lund, Sweden}

\author{Heloisa N. Bordallo}
\affiliation{Helmholtz-Zentrum Berlin f\"ur Materialien und Energie (HZB),  Hahn-Meitner Platz 1, 14109, Berlin, Germany}

\author{Dimitri N. Argyriou}
\affiliation{Helmholtz-Zentrum Berlin f\"ur Materialien und Energie (HZB),  Hahn-Meitner Platz 1, 14109, Berlin, Germany}
\affiliation{European Spallation Source ESS AB, Box 176, 22100, Lund, Sweden}

\author{J. Paul Attfield}
\email[Email of corresponding author:]{j.p.attfield@ed.ac.uk}
\affiliation{School of Chemistry, Joseph Black Building, King's Buildings, West Mains Road, Edinburgh, EH9 3JJ}

\date{\today}

\begin{abstract}
We report a combination of physical property and neutron scattering measurements for polycrystalline samples of the one-dimensional spin chain compound \co. Heat capacity measurements show that an effective S = 1/2 state is found at low temperatures and that magnetic fluctuations persist up to $\sim$6.\Tn. Above \Tn = 6.3 K, measurements of the magnetic susceptibility as a function of T and H show that the nearest neighbour exchange is ferromagnetic. In the ordered state, we have discovered a crossover from a metamagnet with strong fluctuations between 5 K and \Tn \ to a state with a 1/3 magnetisation plateau at 2 $<$ T $<$ 5 K. We use neutron powder diffraction measurements to show that the AFM state has incommensurate long range order and inelastic time of flight neutron scattering to examine the magnetic fluctuations as a function of temperature. Above \Tn, we find two broad bands between 3.5 and 5 meV and thermally activated low energy features which correspond to transitions within these bands. These features show that the excitations are deconfined solitons rather than the static spin reversals predicted for a uniform FM Ising spin chain. Below \Tn, we find a ladder of states due to the confining effect of the internal field. A region of weak confinement below \Tn, but above 5 K, is identified which may correspond to a crossover between 2D and 3D magnetic ordering.
\end{abstract}

\maketitle
\section{\label{sec:level1}INTRODUCTION}
One-dimensional magnetic materials with quantum (S=1/2) spins have been widely studied as their  excitation spectra often consist of continuums of fractionalized particles \cite{alan,nagler} \ which have no analogues in higher spin or three$-$dimensional materials. A particularly elegant example is the Ising spin chain, which in a transverse field undergoes one of the best understood quantum phase transitions\cite{Pfeuty,Sachdev,coldea}. Much of the interesting physics in Ising chains, such as propagation of domain walls (solitons) only occurs when neighbouring chains are weakly coupled together in real materials. For example, the excitations of an isolated antiferromagnetic Ising chain consist of blocks of reversed spins. Intrachain coupling mixes these states together creating a broad band centred in energy at twice the interchain coupling constant\cite{villain}. Thermal population of these states creates a low energy response (the Villain mode) which has a characteristic square root divergence and can be observed by inelastic neutron scattering.
\\Recently, attention has focused on those materials which have structures more complicated than simple linear chains. The zig-zag chain material \nb \ is a notable example, showing a transition from spin-density wave order to non-collinear magnetic order on cooling\cite{Sharf:1979p121,Mitsuda:1994p3568}. Measurements of physical properties\cite{Maartense:1977p93,Hanawa:1994p2706} \ and neutron scattering\cite{Heid:1995p123,Heid:1997p547,Kobayashi:1999p3331,Kobayashi:1999p9908,Kobayashi:2000p024415} \ under magnetic fields also detect a rich phase diagram which includes regions of three-phase coexistence\cite{Weitzel:2000p12146}.
\begin{figure}[tb!]
\begin{center}
\includegraphics[scale=0.75]{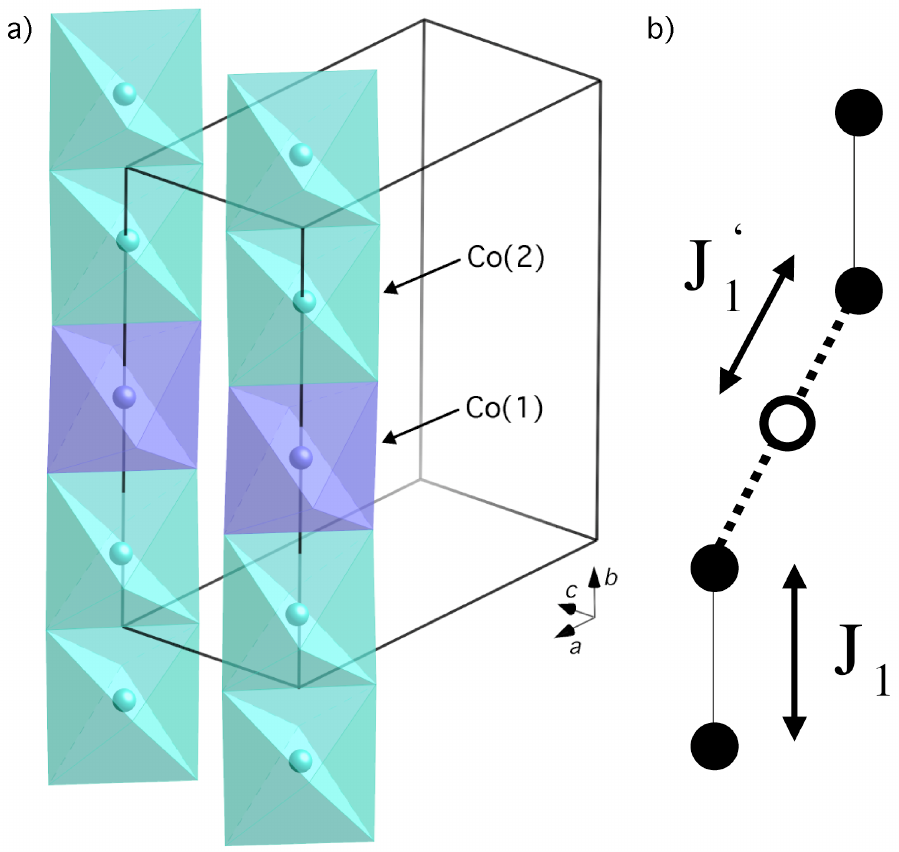}
\caption{(color online) Structure of triclinic \co \ showing edge$-$sharing CoO$_6$ chains, with the two crystallographically independent sites indicated. For clarity, the V$_{2}$O$_{5}$ blocks are omitted. The principle exchange interactions in the chain are indicated in b).}
\label{Fig1}
\end{center}
\end{figure}
\\Further increases in structural complexity can also produce groundstates which are highly sensitive to external pertubations such as magnetic fields. For example, \ca\ has a structure consisting of CoO$_6$ chains on a triangular lattice\cite{ca} and undergoes a counterintuitive magnetic $disordering$ at the lowest temperatures. Applied fields in this material find a cascade of magnetisation plateaus\cite{kag}\  which has been attributed to quantum tunnelling effects similar to those seen in molecular magnets\cite{QT}.
\\In this paper we report our observations on \co, which crystallises in two polymorphs\cite{mono,tri}. Here we study the triclinic ($P\bar1$) brannerite polymorph (the basic properties of the monoclinic polymorph have also recently been reported\cite{jacs}). The structure of \co \ (Fig. 1a) consists of well separated edge-sharing chains of CoO$_6$ octahedra running down the $b$ axis interspersed by non-magnetic V$^{5+}$ in V$_2$O$_5$ like blocks\cite{tri}. The  Co-Co distance within the chains is $\sim$ 3 \AA \  and the distances between chains are $\sim$  4.8 \AA \ along [001] and $\sim$ 7.2 \AA \ along [100]. The chains contain two crystallographically independent sites (Fig. 1b) which alternate in the sequence 1.1.2.1.1 etc. The main magnetic exchange interactions are also shown in Fig. 1b, with a notable modulation along the chain direction (i.e. J$_{1}$ $\neq$ J$_{1}^{'}$). The magnetic properties of several other brannerites have been reported. The isostructural material CuV$_2$O$_6$ is a low dimensional antiferromagnet, due to Cu$^{2+}$ orbital order\cite{Vasil•À?ev:1999p3021}, whilst monoclinic MnV$_2$O$_6$ is an isotropic antiferromagnet that shows reduced magnetic coherence lengths due to Mn/V antisite disorder\cite{me}. 
 \section{\label{sec:level1}EXPERIMENTAL}
 Polycrystalline samples of \co \ were synthesised using a citrate decomposition method\cite{me} \ similar to that used for \mn. Stoichiometric quantities of Cobalt(II) Acetate Tetrahydrate (Aldrich, 99 \%+) and V$_2$O$_5$ (Aldrich, 99.99 \%) were mixed in distilled water together with a three$-$fold molar excess of acetic acid. The mixture was slowly heated and stirred until a gel formed. The gel was allowed to solidify then decomposed at 300$\,^{\circ}\mathrm{C}$ for three hours. The resulting solid was ground, pelleted and heated at 600, 630 and 650$\,^{\circ}\mathrm{C}$ for 12, 12 and 72 hours respectively.  Powder X-ray diffraction using a Bruker D8$-$Advance showed a phase pure product. Magnetic susceptiblity measurements were performed using a Quantum Design MPMS system in varying fields up to 2 T as a function of temperature from 2 - 300 K in field and zero field cooled conditions. Magnetisation isotherms were recorded using the ac-susceptiblity option of a Quantum Design PPMS system at temperatures from 2 - 35 K in fields of up to 9 T. This system was also used for specific heat measurements. For these, an addenda measurement consisting of the sample mount and a small quantity of grease was performed before a small pellet of \co \ was affixed. We used a variety of neutron powder diffractometers to characterise our samples. High resolution data were collected at 2 K with wavelengths of 1.79 and 2.8 \AA \ using the E9 instrument at HZB, Berlin\cite{e9}. A detector bank consisting of 64 $^{3}$He detectors arranged at 2.5 $\,^{\circ}$ \ intervals was employed. A total of 32 steps of the detector were carried out, resulting in a step size of 0.078 $\,^{\circ}$ with a resolution minimum of $\Delta d/d$ $\sim$ 2 x 10$^{-3}$ over a total angular range of 160 $\,^{\circ}$. The high resolution data were refined by the Rietveld method using the GSAS suite of programs\cite{gsas}. We also performed temperature dependent measurements using the medium resolution focussing diffractometer E6, also at HZB\cite{e6}. This instrument has a vertically and horizontally bent pyrolytic graphite monochromator and two large position sensitive detectors giving a high flux at a wavelength of 2.4 \AA. For both diffraction measurements, the sample was held in a 6 mm vanadium can and temperature was controlled using a standard Orange helium flow cryostat. Time of flight inelastic neutron scattering data were collected with  IN5 at the ILL, Grenoble. This instrument is a direct geometry cold neutron spectrometer with a 30 m$^{2}$  array of position sensitive detectors. The sample was contained in an annular aluminium can placed in an orange cryostat. All data were background corrected and normalised to the incoherent scattering of vanadium using standard routines in the program LAMP. We employed two configurations with incident energies of 3.27 and 7.08 meV, giving resolutions of around 0.1 and 0.35 meV respectively.
  \section{\label{sec:level1}RESULTS}
  \subsection{\label{sec:level2}Physical Properties}
  The magnetic susceptibility of \co, measured in a 500 Oe field, is shown in Fig. 2a and shows a sharp transition to an antiferromagnetically ordered state below \Tn \ = 7 K. No divergence between field and zero field cooled measurements was seen. The inverse susceptibility of \co \ is well fitted by the Curie-Weiss law in the range 125 $-$ 300 K with a fitted moment of 5.22(2) \MuB \ and a Weiss temperature of 9.2(2) K. The moment shows a substantial increase due to spin orbit coupling from the spin only value for S = 3/2 (3.87 \MuB) as is often found for high spin Co(II) in octahedral coordination with a 4T$_{1g}$ ground state. The plot of $\chi$.T vs. T shown in the inset of Fig. 2a, provides strong evidence for nearest neighbour ferromagnetic correlations, as a sizeable enhancement in $\chi$.T is observed below 40 K, before an overall antiferromagnetic state is realised at \Tn. These strong ferromagnetic correlations presumably originates in the CoO$_6$ chains which have the shortest exchange pathway by far. Long range order is therefore the result of antiferromagnetic interchain (or plane) exchange.
\begin{figure}[tb!]
\begin{center}
\includegraphics[scale=0.55]{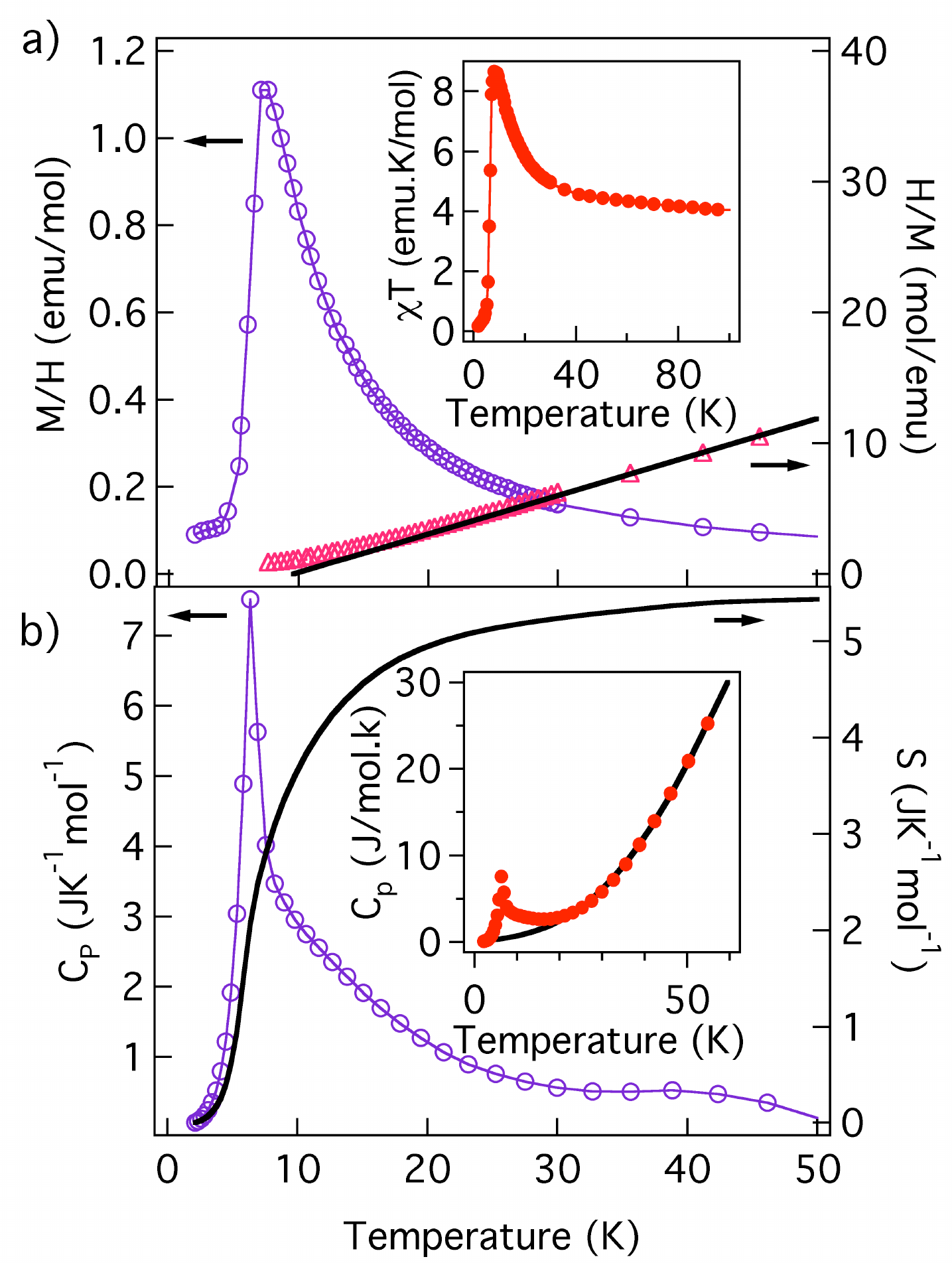}
\caption{(color online) (a) Magnetic susceptibility and inverse susceptibility of \co, the black line shows the Curie-Weiss fit to high temperature region described in text. Inset shows $\chi$.T as a function of temperature. (b) Magnetic specific heat of \co, line shows integrated magnetic entropy, inset shows total specific heat and estimated lattice contribution.}
\label{Fig1}
\end{center}
\end{figure}
\begin{figure}[tb!]
\begin{center}
\includegraphics[scale=0.55]{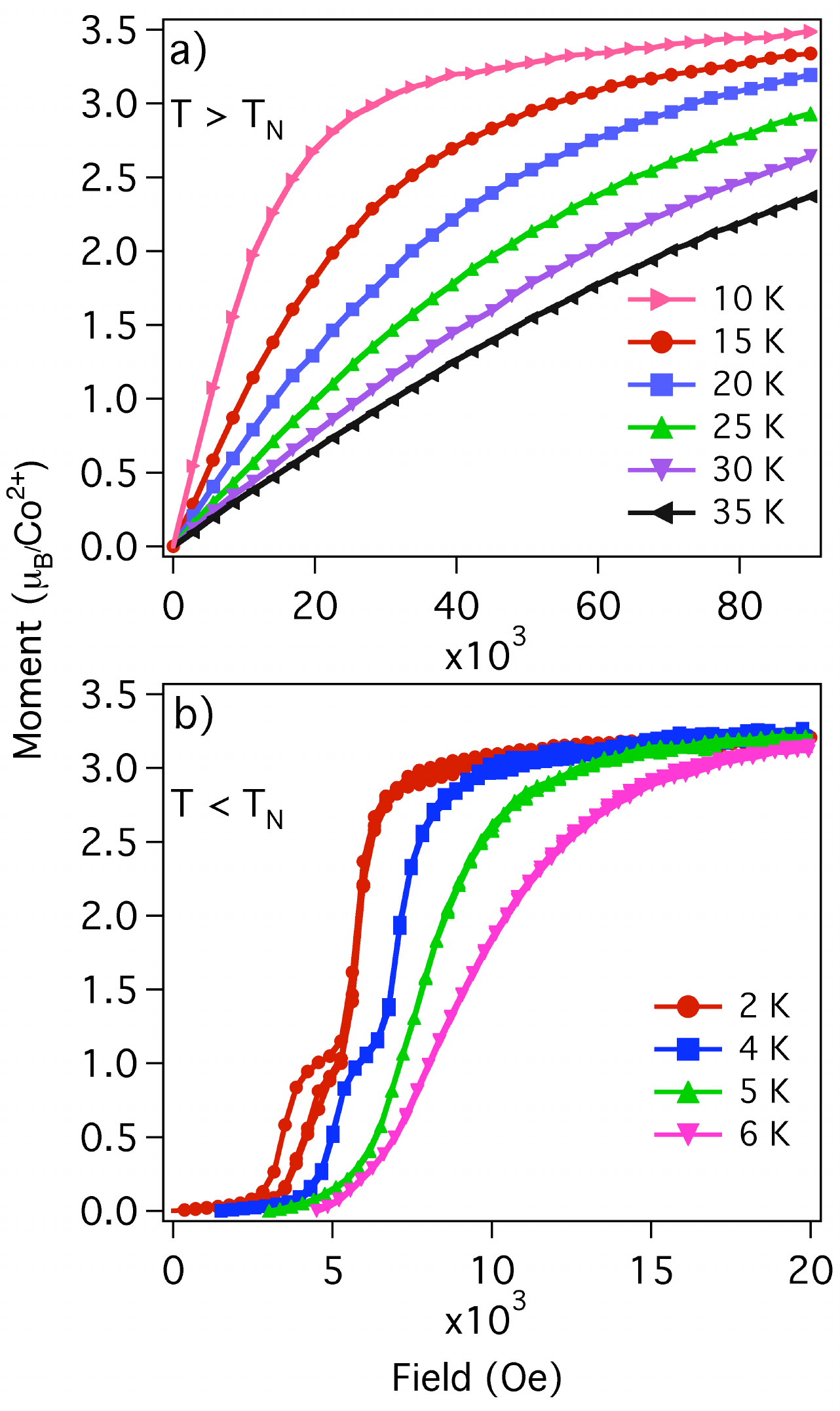}
\caption{(color online) (a) Magnetisation isotherms measured for \co \ above magnetic ordering temperature; (b) Magnetisation isotherms measured for \co \ below magnetic ordering temperature. A constant $x$-axis offset of 1500 Oe has been applied to the results in (b) for clarity. Lines in all plots are guides to the eye.}
\label{Fig1}
\end{center}
\end{figure}
  \\The measured specific heat of a polycrystalline pellet between 2 and 70 K is shown in Fig. 2b. A sharp lambda anomaly is seen at 6.3 K, showing that only one magnetic transition occurs in this temperature range. Application of a 1 T field (which is above the metamagnetic field) both suppresses and broadens the peak (not shown). Differences between the specific heat in zero and 1 T applied field were found up to 30 K, which suggests that magnetic fluctuations persist to at least this temperature, in agreement with the magnetic measurements reported above. We estimated the lattice contribution to the specific heat of \co \ by fitting the high temperature (T $>$ 30 K) specific heat to a function of the form $A$.T$^3$ + $B$.T$^5$, the results of which are shown in Fig. 2b as a bold line. The magnetic entropy was obtained by subtracting the lattice contribution, the total obtained (5.43 J/mol.K) is very close to the expected value for an S = 1/2 spin (R.ln2 = 5.76 J/mol.K). This result implies that the groundstate of the Co$^{2+}$ cations is well described by a Kramers doublet, and that strong single ion anisotropy will dominate the magnetic properties in the temperature region studied here. We note that 65 \% of the entropy is obtained above \Tn, which confirms the low dimensional nature of \co.
\\M(H) isotherms were recorded at a range of temperatures in the magnetically ordered region and above \Tn \ in fields of up to 9 T. Above \Tn, the M(H) isotherms were characteristic of a soft ferromagnet (Fig. 3a). These observations corroborate the evidence from the susceptibility measurements for nearest-neighbour ferromagnetic exchange in \co. Based on the isotherms shown in Fig. 3a, these correlations can be seen to extend well above \Tn, to $\sim$35 K. Isotherms measured in the range 2 $<$ T $<$ \Tn \ are shown in Fig. 3b. At 2 K, two field induced transitions are seen. The first transition at 3600 Oe corresponds to a plateau at almost exactly one third of the saturation magnetisation (0.95 \MuB), the second transition at 5900 Oe is a metamagnetic transition to a plateau that shows the full saturation magnetisation, 2.9 \MuB \ at 9 T ($\mu_{sat}$ = 2.S  = 3\MuB). Considerable hysteresis was observed around the 1/3 plateau at the lowest temperatures measured. M(H) isotherms at higher temperatures show that the 1/3 plateau becomes non-hysteretic at 4 K and disappears between 4.5 and 5 K (Fig. 3b) whilst the metamagnetic transition persists up to the N\'{e}el temperature.
\subsection{\label{sec:level2}Neutron powder diffraction measurements}
Our neutron powder diffraction measurements at 2 K showed the presence of magnetic order and were also used to refine the crystallographic structure. The data were analysed as follows. First, the crystal structure of \co \ was refined from the short wavelength (1.7973 \AA) data collected on the high resolution diffractometer E9. Due to the low symmetry, the lattice parameters and background were first refined using a Le Bail parameter-less fit, giving results similar to those reported at room temperature. The peak shape parameters (a pseudo$-$Voigt function), scale function and detector zero point shift were then refined before the internal coordinates were allowed to vary. In order to reduce the number of variables, and because of the low scattering length of vanadium, the atomic displacement parameters (ADP) of the Co and V sites were constrained to be equal, as were the oxygen ADP's.
\begin{figure}[tb!]
\begin{center}
\includegraphics[scale=0.3]{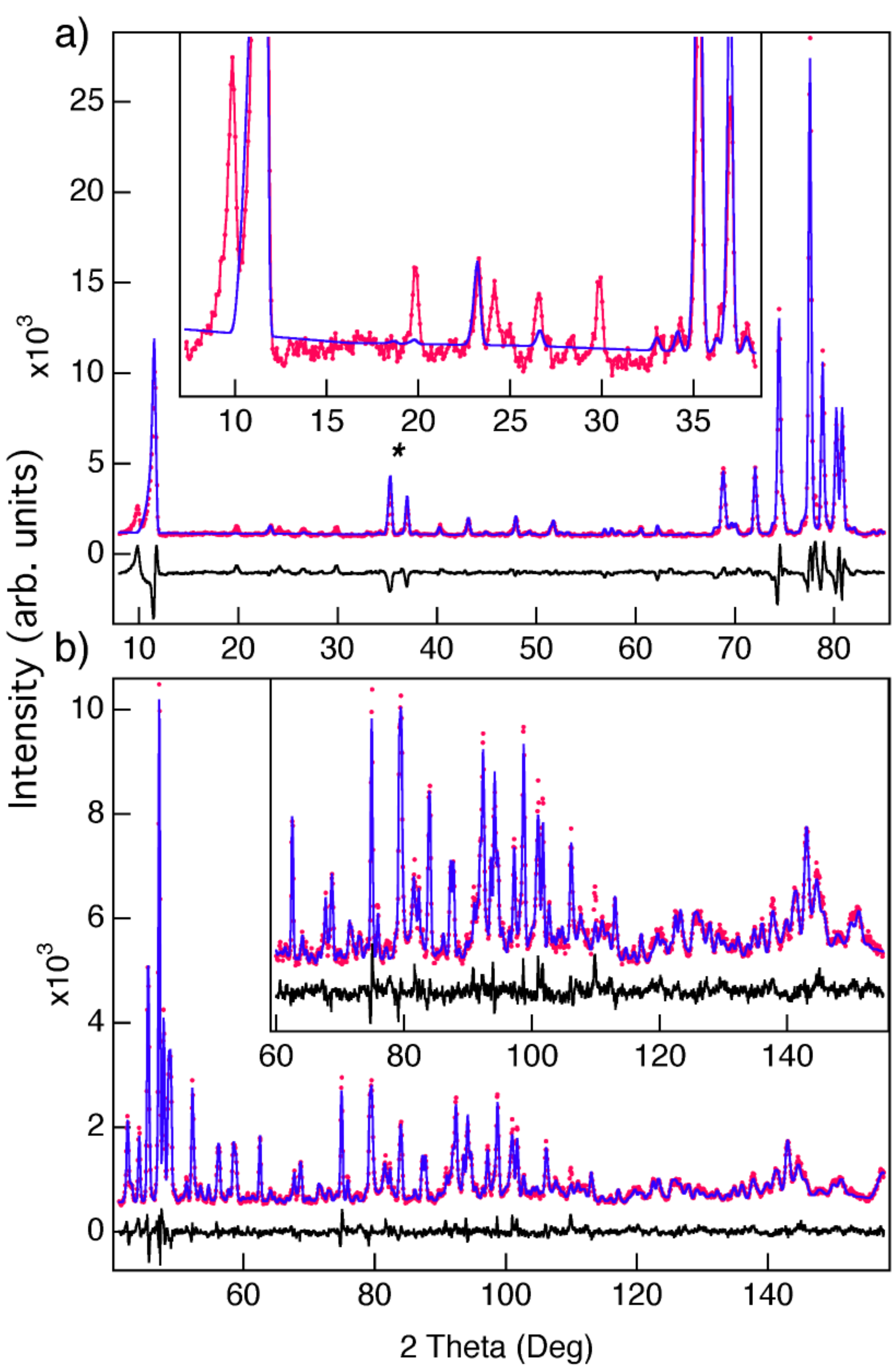}
\caption{(color online) Observed, calculated and difference plots for Rietveld fit to a) 2.8 \AA \ and b) 1.79 \AA \ neutron powder diffraction profiles of \co \ at 2 K. Insets show unindexed magnetic satellite reflections and fit to high angle data respectively. The star in a) marks the magnetic reflections followed to determine the order parameter using E6}
\label{Fig1}
\end{center}
\end{figure}
\begin{table}
\caption{\label{tab:table2}Refined atomic coordinates for \co \ at 2 K from neutron powder diffraction. Lattice parameters were $a$ = 7.172(1), $b$ = 8.875(1), $c$ = 4.8088(8) \AA, $\alpha$ = 90.288(2), $\beta$ =93.931(2), $\gamma$ = 102.160(1) Deg. Refined ADP's are Co/V = 0.2(1), O = 0.73(3) \AA$^{2}\times10^{2}$}
\begin{ruledtabular}
\begin{tabular}{cccc}
 $Atom$ &$x$ &$y$ &$z$\\
\hline
  Co(1)& 0 & 0.5 & 0 \\
  Co(2)& 0.0284(29)&    0.1715(25)&    0.028(4)  \\
  V(1)&       0.704(16)&     0.951(16)&     0.462(24) \\
  V(2)&       0.719(17)&    0.631(15)&     0.433(24) \\
  V(3)&       0.498(16)&     0.240(14)&     0.143(21) \\
  O(1)&       0.1620(14)&    0.4970(11)&    0.3456(19) \\
  O(2)&       0.8413(15)&    0.6399(13)&    0.1623(21) \\
  O(3)&       0.1746(13)&    0.6964(13)&    0.8841(19) \\
  O(4)&       0.1537(13)&    0.0250(12)&    0.8270(19) \\
  O(5)&       0.1697(14)&    0.8925(11)&    0.3428(21) \\
  O(6)&       0.7877(12)&    0.8019(14)&    0.6320(18)  \\
  O(7)&       0.4748(14)&    0.9122(12)&    0.7127(20)  \\
  O(8)&       0.4724(14)&    0.5803(12)&    0.7046(21) \\
  O(9)&       0.5222(13)&    0.7539(14)&    0.2123(18) \\
    \end{tabular}
\end{ruledtabular}
\end{table}
\begin{table}
\caption{\label{tab:table2}Selected refined distances and angles for \co \ at 1.6 K from neutron powder diffraction.}
\begin{ruledtabular}
\begin{tabular}{cc}
 &Bond Distance/Angle (\AA/Deg)\\
\hline
Co(1)$-$O(1) x 2&          1.962(9) \\
Co(1)$-$O(2) x 2&          2.036(11) \\
Co(1)$-$O(3) x 2&         2.024(11) \\
 \\
Co(2)$-$O(2)&          1.993(25) \\
   Co(2)$-$O(3)&         2.113(25) \\
   Co(2)$-$O(4)&        2.005(23) \\
 Co(2)$-$O(4)&         2.099(25) \\
   Co(2)$-$O(5)&        2.193(21)\\
   Co(2)$-$O(6)&        2.009(21)\\
   \\
   Co(1)$-$O(2)$-$Co(2)&           94.8(8) \\
    Co(1)$-$O(3)$-$Co(2)&          91.6(7)\\
   Co(2)$-$O(4)$-$Co(2)&           93.5(9)\\
\end{tabular}
\end{ruledtabular}
\end{table}
 The possibility of antisite disorder between Co and V was investigated, but failed to give an improvement in the fit residuals and was therefore discounted. The refinement for this histogram converged with R$_{wp}$ = 0.0664 and R$_{p}$ = 0.0512 and gave the results shown in Tables I and II. Notably, the refined Co$-$O$-$Co bond angles support our suggestion of ferromagnetic superexchange interactions along the chain direction, as in all cases, these are less than $\sim$95 $\,^{\circ}$. The observed, calculated and difference plots for the Rietveld fits are shown in Figs 4a and 4b. 
 \\At low angles, magnetic reflections which did not index on the chemical unit cell were observed (Fig. 4a). Previous work on related Mn- and CuV$_{2}$O$_{6}$ materials\cite{me,cumag} \ has shown that simple collinear groundstates consisting of ferromagnetic chains coupled antiferromagnetically are found. However, neither of these compounds has strong single ion anisotropy which is expected to complicate the magnetic ordering. In the case of \co, we were able to index the largest magnetic reflections on a supercell with $k$=(1/2,0,0). A simple magnetic structure of ferromagnetic planes coupled antiferromagnetically accounted for the majority of the magnetic scattering. However, we also detected a large number of extra satellite reflections (see insets to Fig. 4a). These resisted all attempts at indexing due to the low crystallographic symmetry and probably correspond to a long wavelength modulation of the magnetic structure as seen\cite{camag} \ in e.g. \ca. Incommensurate magnetic structures, often with multiple propagation vectors, are common in systems with competing interactions. In this case, interplay between superexchange and single ion anisotropy on the two crystallographically independent sites is a possible origin. Future single crystal diffraction studies will be necessary to solve this structure. 
 \\ In order to determine the extent of critical fluctuations in \co, we measured the intensity of several fundamental magnetic reflections as a function of temperature using E6. We chose the reflections between 30 and 40 deg marked in Fig. 4a. A great deal of diffuse scattering was observed around the most intense magnetic reflection ($\sim$11 Deg.) near \Tn, precluding use of this region of the diffraction patterns. A plot of the resulting normalised magnetic intensity versus reduced temperature is shown in Fig. 5 (assuming a N\'{e}el temperature of 6.3 K) A crossover between critical behaviour and true long range order is clearly seen at $\sim$5.1 K, which matches the region identified by the M(H) isotherms without a 1/3 magnetisation plateau. A fit of a power law function gave a critical exponent, $\beta$ = 0.22(2), which is at the upper end of the range found for anisotropic layered Ising systems\cite{beta}, although strongly reduced from the value\cite{thesis} \ of 0.35 found for MnV$_{2}$O$_{6}$. Future measurements with better temperature resolution will be needed to determine a more accurate value.
 \begin{figure}[tb!]
\begin{center}
\includegraphics[scale=0.7]{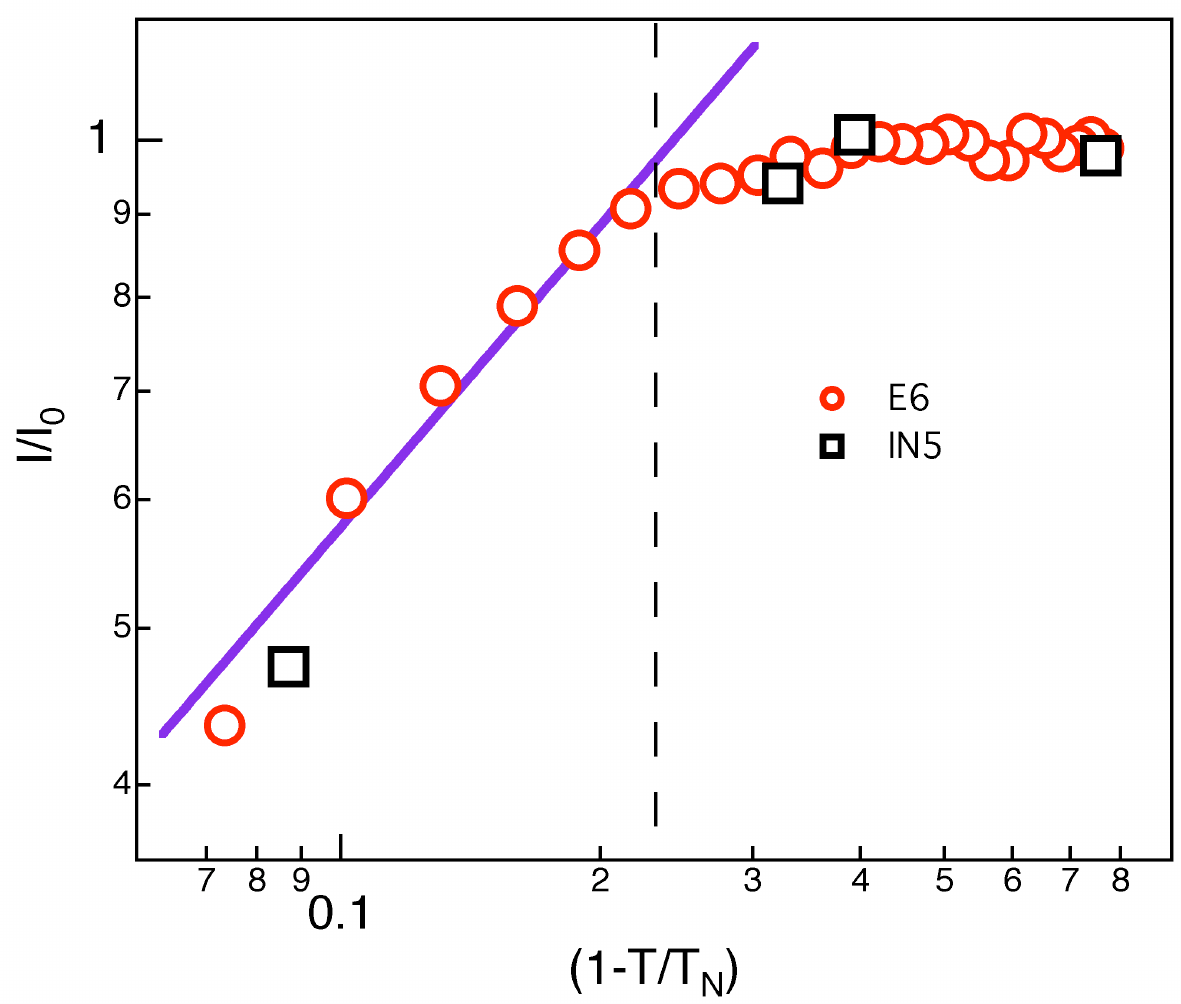}
\caption{(color online)Temperature dependance of the elastic magnetic diffraction peaks recorded using E6. The black squares are elastic data recorded using IN5 with 5 \AA \ incident neutrons. The solid line marks a power law fit,  and the dashed line marks the approximate extent of the critical region.}
\label{Fig1}
\end{center}
\end{figure}
 \subsection{\label{sec:level2}Inelastic neutron scattering measurements}
 A summary of the measurements performed with $E_{i}$ = 7.08 meV and T $>$ \ \Tn \ is shown in Fig. 6. Here the excitations are expected to be representative of the one-dimensional chains. As expected from the single ion anisotropy, the majority of the inelastic spectral weight  is gapped at 7 K. Due to the almost non-dispersive nature of the magnetic response, the data are integrated over all momentum transfers and hence approximate to the magnetic density of states. The sharpness of the gapped peaks is significant, as Van Hove singularities are expected in the magnetic density of states in low dimensional materials. This reinforces our observations from the specific heat measurements which showed that only 35 \% \ of the magnetic entropy is recovered at \Tn, and we conclude that \co \ is highly one-dimensional. Next, by following the temperature dependance of the inelastic response, we confirm the magnetic nature of all these features and show the origin of those above and below 1.5 meV is rather different.
\\In the inset to Fig. 6 we show the integrated inelastic response above and below 1.5 meV. On warming, the intensity of the $\sim$1 meV features increases at the expense of the gapped bands up to at least 35 K, before collapsing. We fitted the data in this region to the gap function:
\begin{equation}
I\propto exp(-E_{g}/k_{B}T)
\end{equation}
The fitted gap value of 3.6(3) meV shows that this feature originates in neutrons scattering from thermally populated states in the gapped bands and that these are therefore soliton continuums rather than well defined spin wave excitations. The magnetic nature of the low energy feature is confirmed by the plateau in intensity above 35 K. The drop in intensity at 50 K is caused by the red-shift of the signal into the region below that used for the integration (i.e. the elastic line) and hence is not intrinsic. Structure also develops in the gapped bands on warming (principally a small increase in intensity at 3.8 meV). Future studies will be needed to determine if this is also domain wall scattering, or marks a cross-over to a more spin-wave like response as more crystal field levels are populated at higher temperatures. At temperatures far in excess of the magnetic exchange scale (200 K), the inelastic response disappears to be replaced by quasielastic scattering showing that no phonons are present in this energy and momentum range (not shown).
\begin{figure}[tb!]
\begin{center}
\includegraphics[scale=0.7]{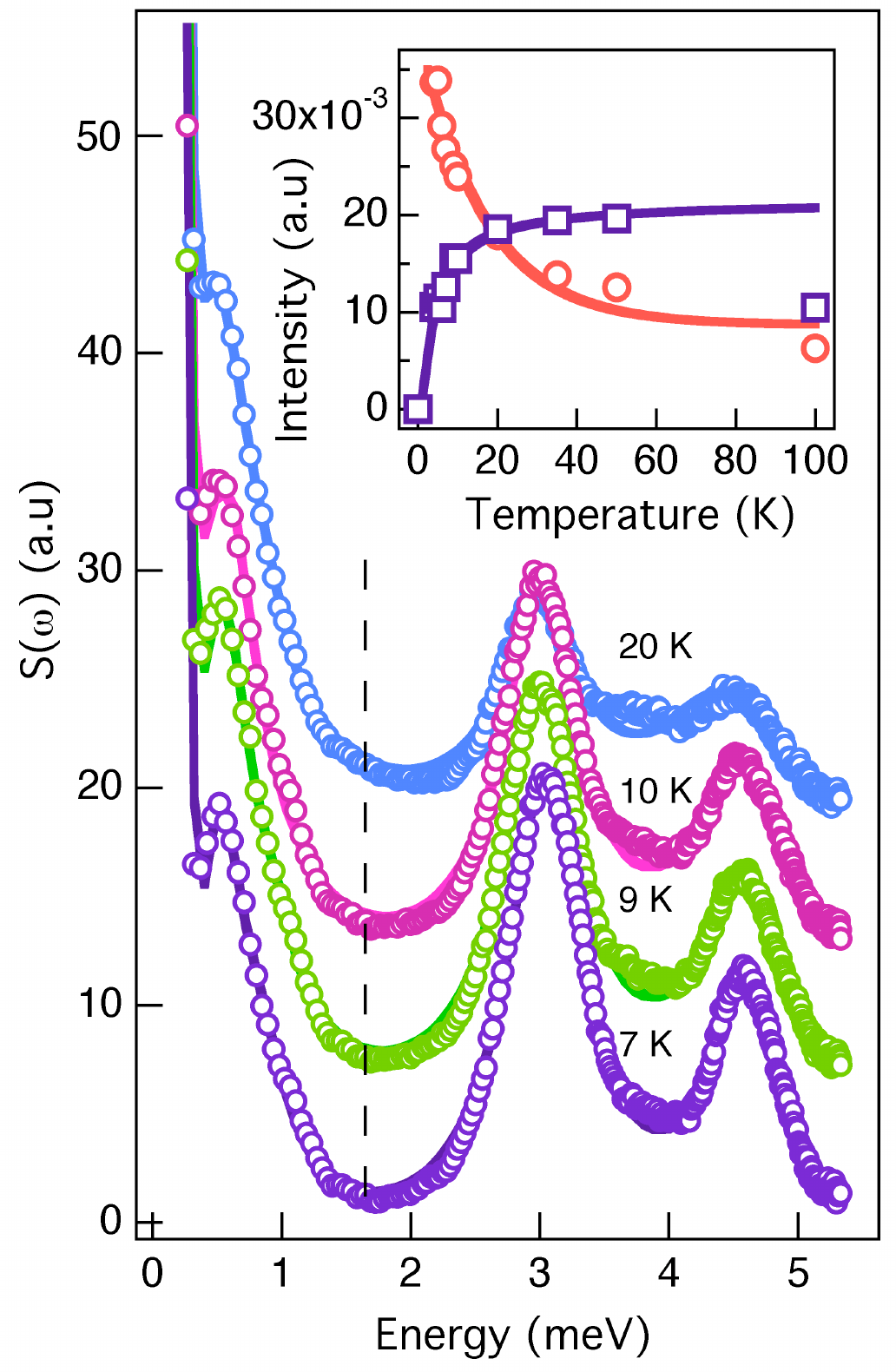}
\caption{(color online) Inelastic neutron scattering from \co \ for 7 $<$ T $<$ 20 K. Date have been summed for all momentum transfers. The inset show the temperature dependence of the integrated inelastic scattering above and below 1.5 meV. All lines are fits as described in the text.} 
\label{fig2}
\end{center}
\end{figure}
\\We next examine in more detail the features at 7 K, including their energy (Fig. 7a)  and Q-dependencies (Fig. 7b). The energy of the peaks shown in Fig. 7a, extracted using the instrument resolution function and Lorentzian peak shapes was 3.05(1) and 4.60(1) meV. For comparison,  the asymmetric feature at low energy seen in Fig. 6 was peaked at 0.47(5) meV (fit not shown). In Fig 7b, the integral of the inelastic response for all energies (i.e. S(Q)) is shown. This approximately follows the squared Co$^{2+}$ form factor (blue line) as expected for nearest$-$neighbour FM exchange.  The individual Q-dependencies of the two features in S($\omega$) are in fact rather different (inset, Fig. 7b). This helps to show that these are spin rather than  crystal field excitations, which would follow the form factor. Crystal field excitations are also generally found at much higher energies in Co$^{2+}$ containing oxides \ \cite{cfield}. Taking into consideration the peculiar dimer arrangement of Co(2) sites in \co \ (Fig. 1a), and the peaked nature of the 4.60(1)  meV feature, we performed an additional test against nearest$-$neighbour AFM coupling. We used the following powder averaged form factor:
\begin{equation}
S(Q)\propto f(Q)^{2}.(1-\sin(Q.d)/Q.d)
\end{equation}
\begin{figure}[tb!]
\begin{center}
\includegraphics[scale=0.15]{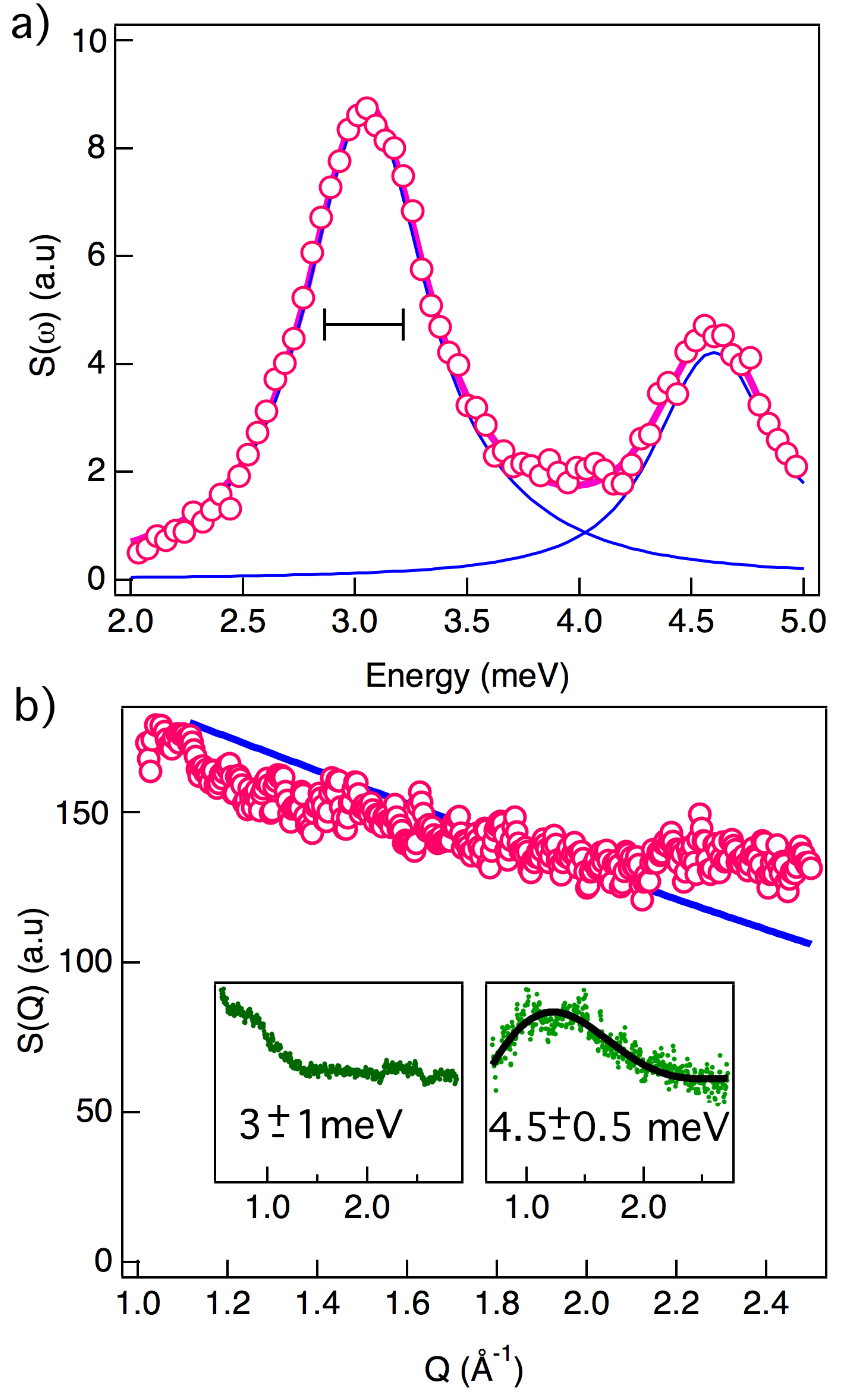}
\caption{(color online) a) Inelastic neutron scattering from \co \ just above \Tn \ at 7 K. Date have been summed for all momentum transfers, the lines are fits as discussed in the text, the horizontal bar shows the instrumental resolution; b) Energy integrated inelastic neutron scattering data for \co, the solid line shows the squared Co$^{2+}$ form factor.  The insets to b) show the Q-dependencies of the two principle excitations.} 
\label{fig2}
\end{center}
\end{figure}
Here $f$(Q) is the magnetic form factor and $d$ a characteristic distance. We also included a background term which accounts for the slow increase in intensity with Q found throughout our data. This is probably due to multiple incoherent scattering of phonons. The fitted value of $d$ was 3.36(2) \AA, which far exceeds the real value in the chains ($\sim$ 2.85 \AA). This shows that \co \ is not a loosely linked chain of dimer units and that the origins of the two features lie elsewhere.
\\ To better define the thermally populated inelastic scattering below 1.5 meV, we show the results of the measurements in the low energy configuration of IN5 over a wide range in temperature in Fig. 8. To remove the intrinsic effects of temperature, we have employed the fluctuation dissipation theorem to extract the generalized susceptibility  $\chi$''(Q,$\omega$):
\begin{equation}
S(Q,\omega)\propto\frac{\chi''(Q,\omega)}{1-exp(-\hbar\omega/k_{B}T)}
\end{equation}
The low energy signal is clearly resolved into two peaks just above  \Tn \ as shown by the  cuts at Q = 1.3 $\pm$0.1 \AA$^{-1}$ in Fig. 8. 
\begin{figure}[tb!]
\begin{center}
\includegraphics[scale=0.55]{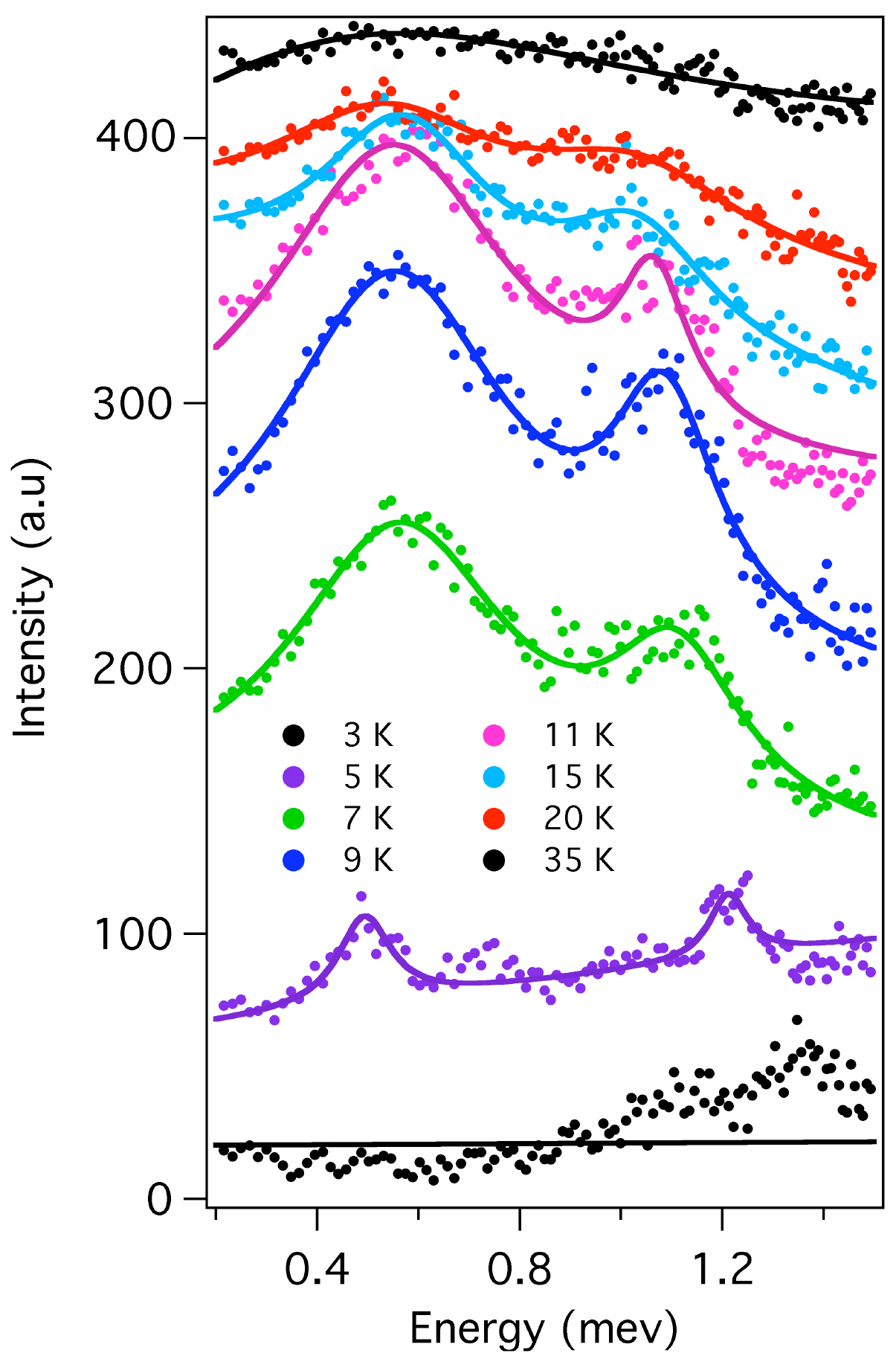}
\caption{(color online) The results of the inelastic neutron scattering experiment using the low energy configuration of IN5 are shown for a range of temperatures. The data are presented in the form of constant Q = 1.3 $\pm$0.1 \AA$^{-1}$ and have been corrected for the effects of temperature using equation (3). The lines show the results of fitting Lorentzian functions and a sloping background. } 
\label{fig2}
\end{center}
\end{figure}
This response reaches a maximum at $\sim$11 K and collapses with further heating. In this region the data were fitted by the sum of two Lorentzians and a sloping background. The position of the excitations was 0.55(1) and 1.06(1) meV at 11 K. Upon heating,  a gradual broadening and convergence of the two peaks was observed. Above 20 K, the data could be equivalently described by a single damped peak. These low energy features thus approximately match the temperature region where our magnetic susceptibility and specific heat measurements detect low$-$dimensional ferromagnetic fluctuations (Fig. 2).
\begin{figure}[tb!]
\begin{center}
\includegraphics[scale=0.09]{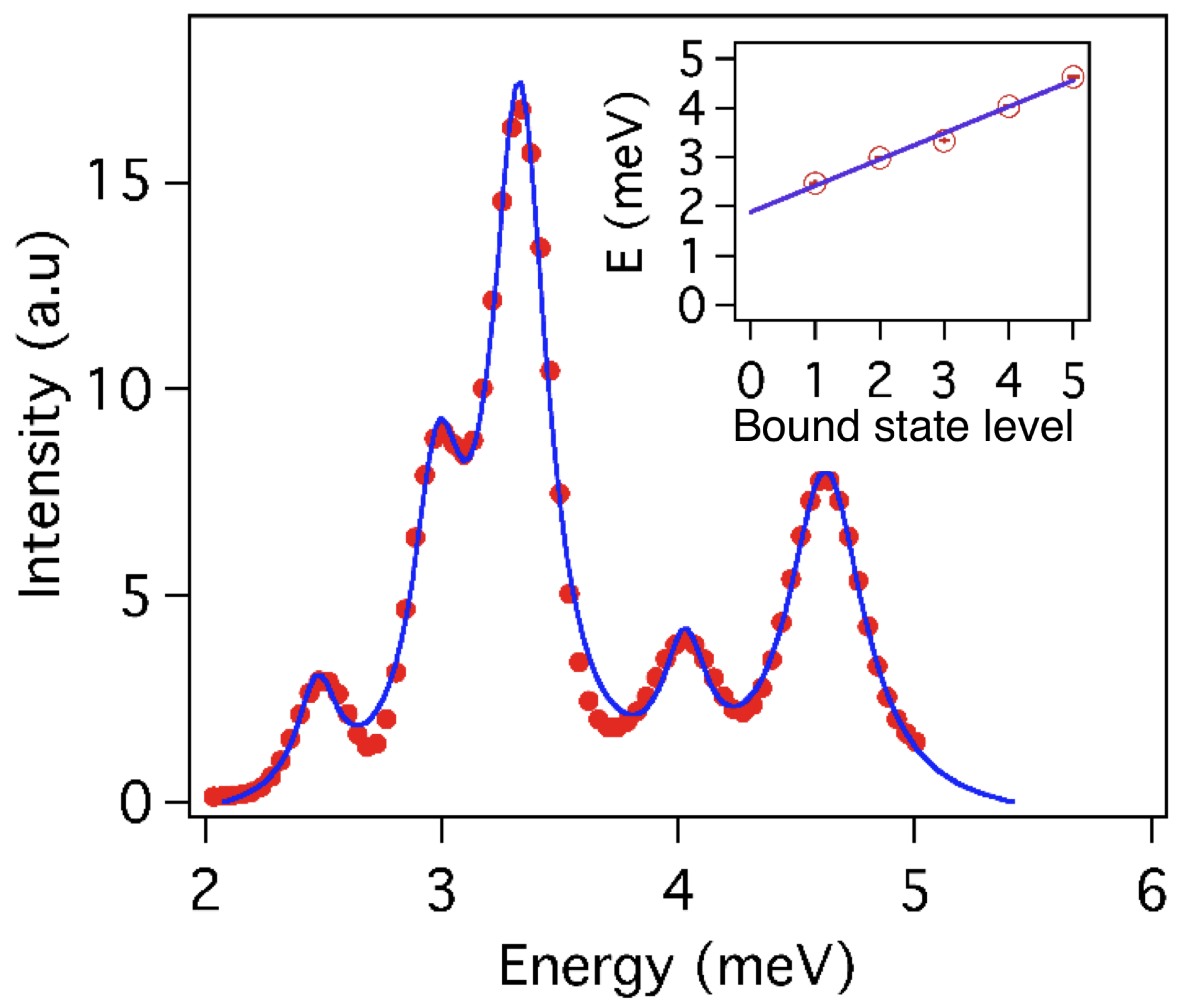}
\caption{(color online) Momentum transfer summed inelastic scattering of \co \ at 1.5 K and 7.08 meV incident energy showing the confining effect of the internal field in the magnetically ordered state. Lines are fits to Lorentzian functions. The inset shows the energy of the bound states with a linear fit.} 
\label{fig4}
\end{center}
\end{figure}
\\The onset of magnetic order, and the associated internal field in \co \ dramatically changes both the low and high energy spin excitations. As shown in Fig. 8, the former almost entirely disappear below 5 K. The gapped features are also strongly affected, as shown by the data collected in the higher energy configuration. Figure 9 shows that, below \Tn, a more complex structure emerges from the broad bands at 3 and 4.5 meV.  The two bands of excitations found above T$_{N}$ split into at least five distinct modes, whose fitted positions (extracted using Lorentzian functions) are shown in the inset to Fig. 9.  Assuming these are bound states, as discussed more fully below, an approximately linear increase in energy with the bound state level was found (inset, Fig. 9). 
\subsection{\label{sec:level1}DISCUSSION}
The above results represent a comprehensive investigation of the structure, magnetic properties and dynamics of \co. Our principal conclusion is that  \co \ may be regarded as a one-dimensional spin chain with considerable Ising type anisotropy and ferromagnetic nearest neighbour exchange. This is supported by both physical property and neutron scattering measurements. Here, we will briefly review the former, before commenting in detail on the latter and the phase diagram which we derive for \co. Turning first to the physical property measurements in low magnetic fields, both the magnetic susceptibility (Fig. 2a) and specific heat of \co \  (Fig. 2b) show a single antiferromagnetic transition at T$_{N}$= 6.3 K. However, the positive Weiss temperature of 9.2(2) K, and the peak seen in $\chi$.T vs. T above T$_{N}$ imply nearest neighbour ferromagnetic exchange. Our specific heat measurements confirm the presence of magnetic fluctuations up to temperatures of around 40 K. These measurements also provide the first piece of evidence for Ising anisotropy as the integrated magnetic entropy is very similar to that expected for an effective S = 1/2 spin. This picture is strongly confirmed by the M(H) isotherms shown in Figs. 3a and 3b and the critical exponent $\beta$ = 0.22(2) derived from neutron powder diffraction (Fig. 5). These physical property measurements compare well with those previously reported \cite{Schneider,Weitzel} \ for Ising FM salts such as CoCl$_{2}$.2H$_{2}$O, which also show a 1/3 magnetisation plateau. \\
Our results start to differ from those reported for these simple materials, as we consider our neutron scattering data. Firstly, our diffraction measurements at 2 K show that the magnetic order is incommensurate, although the majority of the scattering can be modelled assuming FM chains coupling antiferromagnetically (Fig. 4). Secondly, as we now discuss in more detail, our inelastic neutron scattering results do not coincide with predictions for a linear FM Ising spin chain. Unlike AFM Ising chains, soliton excitations in materials with a FM nearest$-$neighbour exchange J$_{1}$, do not deconfine. The excitations are instead isolated flips of blocks of spins\cite{villain} and the results of zero field inelastic neutron scattering may be modelled using standard spin wave theory\cite{inela}. The degeneracy of these excitations may also be lifted by applying magnetic fields, and the resulting spectrum of spin flips has also been observed using infrared transmission experiments\cite{infra}. The observation of sharp, almost dispersionless features in \co, as well as temperature induced scattering is inconsistent with these expectations. This shows that any Hamiltonian proposed for \co \ must include terms which account for the low crystallographic symmetry. Of obvious importance here is the period three variation of crystallographically distinct Co sites (Fig. 1). This may result in a variation in the local easy axis on moving along the chain. In a simple model based on band folding, the presence of two sites would also result in two gapped soliton bands as experimentally observed. Furthermore, this may also be consistent with the presence of two low energy features. The fitted gap function for these shows that they originate in Villain mode type scattering in the gapped bands. As both show the same temperature dependence, the doubling may hence be an indication of both intra and inter band scattering.\\
On cooling below T$_{N}$, the complex features which emerge from the gapped bands between 2 and 5 meV are similar to what was reported for the related compound \nb. This is proposed to be a result of the confining effect of the internal field caused by magnetic order on domain walls. Interchain coupling imposes a linearly increasing potential between propagating domain walls, quantising the continuum into a ladder of bound states\cite{coldea,Goff} \  above the energy required to create an isolated spin flip (2.J$_{1}$) . The overall picture seen in \co \ (inset, Fig. 9) qualitatively agrees with this picture, however significant dispersion is not evident from our data. In Ref. 5, an empirical parameter which tuned the dispersion was introduced into the Hamiltonian. This is strongly dependent on the exchange topology in each individual material. In our powder averaged case, the energies of the bound states increase linearly, and from the $y-$intercept we estimate an average intrachain interaction of 0.95(1) meV.\\ 
The phase diagram of \co \ that we have determined from our physical property measurements is shown in Fig. 10. Of particular note is the 1/3 magnetisation plateau. This feature is not found for all temperatures within the antiferromagnetically ordered region below 6.3 K, instead setting in below $\sim$ 5 K. Interestingly, the region without a plateau coincides extremely well with the region of critical fluctuations identified by neutron powder diffraction shown in Fig. 5. Examination of Fig. 8 also shows that a small amount of the thermally populated domain wall inelastic scattering also remains down to 5 K. This is in some respects similar to the situation found in the intermediate partially disordered magnetic structure in CsCoCl$_{3}$. This compound has a coexistence of 2/3 ordered spins and soliton excitations in a small temperature region before complete order sets in  \cite{pdafm}. A plausible explanation of our results for \co,  consistent with the anisotropic coupling between chains, would be a crossover between 2D and 3D magnetic order below 5 K. 
\begin{figure}[tb!]
\begin{center}
\includegraphics[scale=2.5]{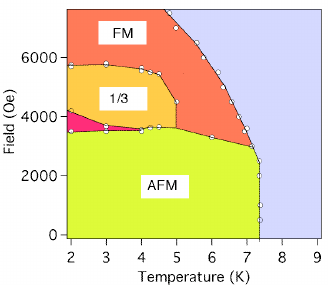}
\caption{(color online) The temperature-magnetic field phase diagram of CoV$_{2}$O$_{6}$ established from the physical property measurements in this work. The various phases are labelled as "FM" (fully polarised ferromagnetic), "1/3" (plateau in magnetisation at 1/3 of saturation) and "AFM" (antiferromagnetic phase). The region of hysteresis found at the lower bound of the 1/3 plateau is shown in magenta.} 
\label{Fig1}
\end{center}
\end{figure}
\subsection{\label{sec:level1}CONCLUSIONS}
We have reported the bulk synthesis and characterisation of powder samples of \co. We have used a combination of physical property and neutron scattering measurements to show that \co \ is a one-dimensional Ising magnet with ferromagnetic nearest neighbour exchange. In the ordered state, we have discovered a crossover from a metamagnet with strong fluctuations between 5 K and \Tn \ to a state with a 1/3 magnetisation plateau at 2 $<$ T $<$ 5 K. This may arise from a dimensionality crossover from 2D to 3D magnetic ordering. The temperature dependence of the inelastic neutron scattering signal shows that the principle excitations are domain walls and at low temperatures we detect signatures of confinement caused by the internal field. It is likely that the low symmetry of \co, especially the presence of two crystallographically independent sites in the edge sharing chains, strongly influences the magnetic properties.
\\Significant challenges remain, including resolving the exact magnetic structure adopted below 7 K and a complete modelling of the inelastic neutron spectra. We anticipate that these can be addressed once single crystals of \co \ become available.
\subsection{\label{sec:level1}ACKNOWLEDGEMENTS}
S.A.J.K  and J.P.A thank the EPSRC and Leverhulme trust respectively for support. We additionally acknowledge the Helmholtz Zentrum Berlin for funding and the Institute Max von Laue-Paul Langevin for access to their instruments. We thank D.A. Tennant for useful discussions.

\end{document}